\documentclass[preprint]{revtex4}
\usepackage{amsmath}
\usepackage{verbatim}
\usepackage{graphicx}
\newtheorem{lemma}{Lemma}
\begin{document}
\title{Thomas-forbidden particle capture}

\author{John H.\ Carter}
\altaffiliation[Present address: ]{Department of Physics,
Astronomy and Materials Science, Missouri State University,
Springfield, MO 65897}
\affiliation{Department of Physics, University of Arkansas,
Fayetteville, Arkansas 72701}

\author{Michael Lieber}
\affiliation{Department of Physics, University of Arkansas,
Fayetteville, Arkansas 72701}

\begin{abstract}
At high energies, in particle-capture processes between ions and atoms, classical kinematic requirements show that generally double collision Thomas processes dominate. However, for certain mass-ratios these processes are kinematically forbidden. This paper explores the possibility of capture for such processes by triple or higher order collision processes.
\end{abstract}

\maketitle

\section{Introduction}
It has long been known that at high energies, particle capture (also called particle or mass transfer or exchange) is dominated by classical kinematics.  As a fundamental three-body process it has been much studied in both 
experiment and theory \cite{ref,Shakeshaft,Dett, Maple}. In 1927 L.H. Thomas \cite{Thomas} did a theoretical analysis of the experiments of Rutherford, and especially G.H. Henderson \cite{Ruth}, in which energetic alpha 
particles captured an electron as they passed through several different media, emerging as $He^+$ ions.

We denote the generic process $1+(2,3)\rightarrow(1,2)+3,$
and ``high energy'' means that the energy of the incident particle 1 is large compared with the binding energy of particle 2 to either the initial nucleus 3 or the final nucleus 1. Thus, without loss of significance we can consider the 
transferred particle 2 to be initially at rest with respect to the nucleus 3, and we define ``capture'' to mean that 2 emerges from the collision with zero velocity with respect to the scattered incident particle 1, i.e. they move off with 
identical speeds and directions. The nature of the interactions between the particles, generally Coulombic, is not important in our considerations, so we consider only hard contact collisions.

Conservation of overall energy and momentum forbids the capture process to take place with only a single collision (except in the case where 1 and 3 have exactly the same mass and 1 collides head-on with 3). In the case 
Thomas considered, there are two collisions: In the first, particle 1 (the alpha particle) collides with the bound particle 2 (the electron). This collision brings the electron from rest to the speed needed for capture, but its 
direction is wrong. The electron then collides essentially elastically with the nucleus 3 to which it was bound, changing its direction to parallel that of the recoiling particle 1. The scattering angle of particle 1 is called 
the ``Thomas angle'' and is given by $(m_2/m_1)\sin 60^{\circ}$.  In the most studied case, protons on hydrogen, this is 0.472 mrad. We call this scenario ``Thomas process A.''

Logically (but not necessarily physically) two other double collision processes are conceivable. In scenario ``Thomas process B'' the first collision between 1 and 2 brings particle 2 to the necessary final speed. Particle 1 
then collides with particle 3 and has its direction changed to enable capture of 2. In scenario ``Thomas process C'' the first collision is between particles 1 and 3. Recoiling particle 3 then collides with particle 2. These three 
possibilities are shown schematically in Fig. 1. Note that in all three cases, the kinematics are fixed since the number of constraint equations (conservation of momentum and energy for each collision) equals the number of 
variables. However, depending upon the mass ratios, not all of these processes may be allowed. That is, conservation of energy and momentum may forbid one or more of these scenarios --- or, in the cases studied below, all 
three may be forbidden.

In 1987 the senior author of this paper devised a simple diagram to show how the available processes depend on the mass ratios \cite{Lieber}. This diagram, which has come to be known as the ``Lieber diagram,'' is shown 
in Fig. 2.  The symmetry of the diagram about the $45^{\circ}$ line may be attributed to time-reversal symmetry. The curved boundaries are simple rectangular hyperbolas. Related diagrams have been given 
in \cite{Dett, ref2}.

The quantum mechanical picture is more complicated 
\cite{ref,Shakeshaft,Dett, Maple}.  Because of the uncertainty principle 
and the necessity of using wave packets, the momenta of the particles and 
their positions are spread out.  At the high energies considered here, the 
Born approximation might be considered a good approximation.  With Coulomb 
potentials it leads to a differential cross section with a large peak in 
the forward direction.  The total cross section falls off as $E^{-6}$.  
However the second Born term dominates the first at sufficiently high 
energies because it falls off only as $E^{-5.5}$, as shown in the 
dissertation of Drisko \cite{Drisko}.  This unique phenomenon occurs 
because of the Thomas double-collision process, which gives rise to a pole 
in the integrand of the second Born term when the propagator is on shell.  
The singularity gives rise to a second peak in the differential cross 
section at the Thomas angle which emerges from the background when the 
projectile energy is sufficiently large. The Thomas peak has been observed
experimentally \cite{Vogt} in the pure three-body case of electron capture 
in proton-hydrogen scattering. An earlier atomic physics observation 
\cite{Horsdal} utilized capture of one electron from helium in proton 
helium scattering - the scenario of Rutherford, Henderson and Thomas - and 
so is somewhat less clear-cut because of the presence of two electrons.

Even earlier there was an observation of the Thomas peak in a molecular 
collision, as proposed by Bates \cite{Cook}: a proton is captured from a 
methane molecule in the process $p+CH_{4}\rightarrow CH_{3}+H_2^{+}$. 
The Thomas peak occurs at 46 degrees.

The question to be explored in this paper is the behavior of collisions for which all three Thomas double-collision processes are kinematically forbidden.  Is capture possible by more than two collisions? We answer this question in the affirmative. Atomic processes which are Thomas-forbidden are rare because of the particular masses of the electron and proton. However, if we admit muons, the process: $H+\mu\rightarrow e + (\mu p)$, which describes the formation of muonic Hydrogen, can be shown to be Thomas-forbidden.  Molecular processes are simpler to come by, e.g. 
$Na+I_2\rightarrow NaI+I$. These and other processes are discussed below. 

Our study reveals interesting structure in the forbidden regions of the Lieber diagram.  We have not determined the quantum mechanical behavior of the cross section for these processes, which would correspond to third and 
higher order terms in the Born series. In this situation there is no on-shell pole in the second Born term, but may be one in the third or higher term. It is not clear whether such a pole would lead to dominance of this term in the 
high energy cross section.

In section II we briefly review the kinematics of the Thomas double collision processes. This will establish the notation and methods to be used for exploring the forbidden processes.

\newpage
\section{Thomas double-collision kinematics}
It is convenient to work with
the ratios of the masses of particles 1 and 3 to the mass of particle 2: $%
a=m_1/m_2$ and $b=m_3/m_2$.  The mass ratios $a$ and $b$ which lead to
scattering may be found simply by solving the equations of conservation of
energy and momentum at each collision.  It is also convenient to refer to the binary collisions using the number of the particle \underline{not} involved in the collision, the "spectator particle."  We will use parentheses to distinguish between particle numbers and collisions numbers.  So, for example, (3) refers to a collision between particles 1 and 2.  A capture process will be described by a string of numbers enclosed by
parentheses corresponding to the collisions in the order they occur from left to right.  Thus the three double collision processes we have considered, A, B, and C, can be referred to as (31), (32), and (21) respectively.  A sketch of their trajectories is given in figure 1.
\begin{figure}
\caption{The three binary processes.  (Bold line = particle 1, Thin line = particle 2, Dashed line = particle 3)}
\begin{center}
\includegraphics[height=0.55\textheight,width=0.55\textwidth]{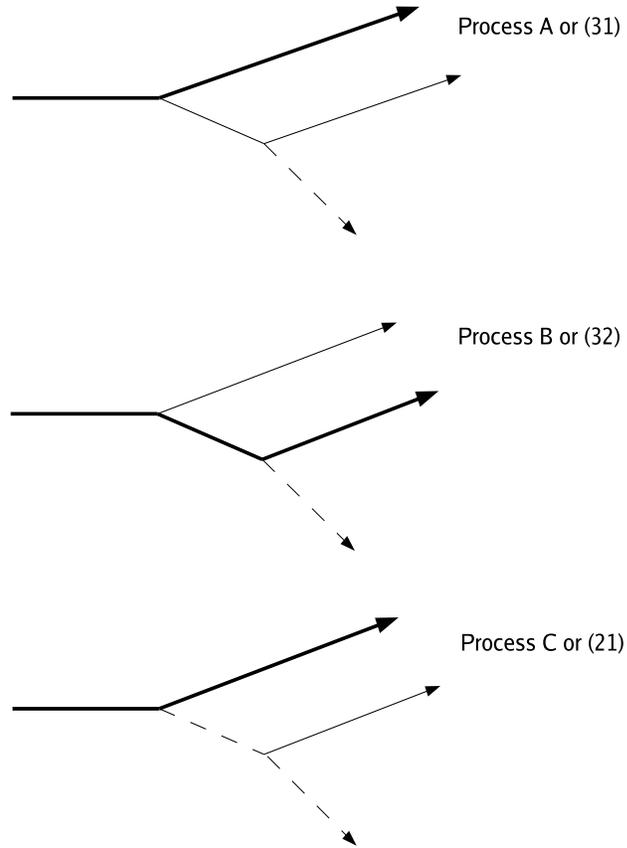}
\end{center}
\end{figure}

We will now determine the mass ratios $a$ and $b$ for which these processes are allowed.

First we look at process A, that is, (31).  The initial velocities of particles 2 and 3 are zero and  $\mathbf{u}$ will be the initial velocity of particle 1.  After the first collision, (3), the velocities of 1 and 2 are $\mathbf{u'}$ and $\mathbf{v}$.  After the second collision, (1), the velocities of 2 and 3 are $\mathbf{u'}$ and $\mathbf{w}$. The equations of conservation of momentum for the two collisions are:
\begin{eqnarray}
a\mathbf{u}&=&a\mathbf{u'}+\mathbf{v}\\
\mathbf{v}&=&\mathbf{u'}+b\mathbf{w}.
\end{eqnarray}

The equations of conservation of energy are:
\begin{eqnarray}
\frac{1}{2}au^2&=&\frac{1}{2}au'^2+\frac{1}{2}v^2\\
\frac{1}{2}v^2&=&\frac{1}{2}u'^2+\frac{1}{2}bw^2.
\end{eqnarray}
While solving these four equations is straightforward, we will use a technique which will make the calculations in Section III simpler.

Without loss of generality we may consider the collisions to occur in a plane, the xy-plane.
If we express the velocities of the three particles as complex numbers and write them as the components of a three dimensional, complex vector, as:  $(v_{1x}+iv_{1y},v_{2x}+iv_{2y}, v_{3x}+iv_{3y})^T,$ then the velocities before and after a collision can be related by a $3\times 3$ complex matrix \cite{Carter}.  If we denote the matrices associated with collisions (1), (2) and (3) by $S_1,$ $S_2,$ and $S_3,$ then we can write:
\begin{eqnarray}
S_1(x)=S_1(0)^{\alpha/\pi}=\frac{1}{1+b}\left(
\begin{array}{ccc}
1+b & 0 & 0 \\
0 & 1+bx & b(1-x) \\
0 & 1-x & b+x
\end{array}
\right)\mbox{,} \\
S_2(x)=S_2(0)^{\alpha/\pi}=\frac{1}{a+b}\left(
\begin{array}{ccc}
a+bx & 0 & b(1-x) \\
0 & a+b & 0 \\
a(1-x) & 0 & b+ax
\end{array}
\right)\mbox{,} \\
S_3(x)=S_3(0)^{\alpha/\pi}=\frac{1}{1+a}\left(
\begin{array}{ccc}
a+x & 1-x & 0 \\
a(1-x) & 1+ax & 0 \\
0 & 0 & 1+a
\end{array}
\right)\mbox{,}  \label{sdef}
\end{eqnarray}
where $\alpha$ is the scattering angle of the collision and
$x=\exp(i\alpha)$.

The matrix which transforms the initial velocities of the three particles to the final velocities is, for process A, $S_1(x_2)S_3(x_1),$ where $x_1=\exp(i\alpha_1)$ and $x_2=\exp(i\alpha_2)$ and $\alpha_1$ and $\alpha_2$ are the scattering angles for the first and second collisions.  The initial three velocities are:  $(u, 0, 0)^T$.  By insisting that after the two collisions, 1 and 2 have the same velocity, we obtain an expression relating $\alpha_1$ and $\alpha_2$:
\begin{equation}
(1,-1,0)S_1(x_2)S_3(x_1)\left(\begin{array}{c}u\\0\\0\end{array}\right)=0.
\end{equation}
Its solution is
\begin{equation}
x_2=\frac{(1+a+b)x_1+ab}{ab(1-x_1)}.
\end{equation}  The condition that $x_2$ have unit absolute value implies that $\cos(\alpha_1)=(ab-(1+a+b))/2ab.$
We need $\cos(\alpha_1)$ to be in the range $[-1,1]$ and this condition is satisfied when $3ab-a-b-1\ge 0.$  Thus the allowed region for process A lies to the right of the rectangular hyperbola given by $3ab-a-b-1=0$.  The equation for the hyperbola can be written as $(a-1/3)(b-1/3)=4/9,$ so the asymptotes are at $a=1/3$ and $b=1/3$.

For process B, or $(32)$, the condition for capture is:
\begin{equation}
(1,-1,0)S_2(x_2)S_3(x_1)\left(\begin{array}{c}u\\0\\0\end{array}\right)=0.
\end{equation}
whose solution is:  $x_2=a(b-(1+a+b)x_1)/b(a+x_1)$.  The absolute value of $x_2$ is unity when $\cos(\alpha_1)=(a^2+ab+a-b)/2ab.$  The cosine falls in the desired range when $3ab+a^2+a-b\ge 0$ and $b\ge a.$  Thus the allowed region for B lies between the line $b=a$ and the curve $3ab+a^2+a-b=0$.  The latter is again a rectangular hyperbola with an asymptote at $a=1/3$ in the first quadrant.

The process C, or $(21)$, is similar to B.  If B is viewed under time reversal and with particles 1 and 3 exchanged, it is equivalent to C.  Thus the values of $(a,b)$ for which C is allowed are simply those of B with the roles of $a$ and $b$ reversed.  Thus C is allowed if $(a,b)$ lies in the region between the the line $b=a$ and the curve $3ab+b^2+b-a=0$.  The latter is a rectangular hyperbola with an asymptote at $b=1/3$.

The allowed regions for the three binary processes are shown in the ``Lieber diagram,'' figure 2.

\begin{figure}
\caption{The Lieber diagram for double collision processes}
\begin{center}
\includegraphics[height=0.38\textheight,width=0.55\textwidth]{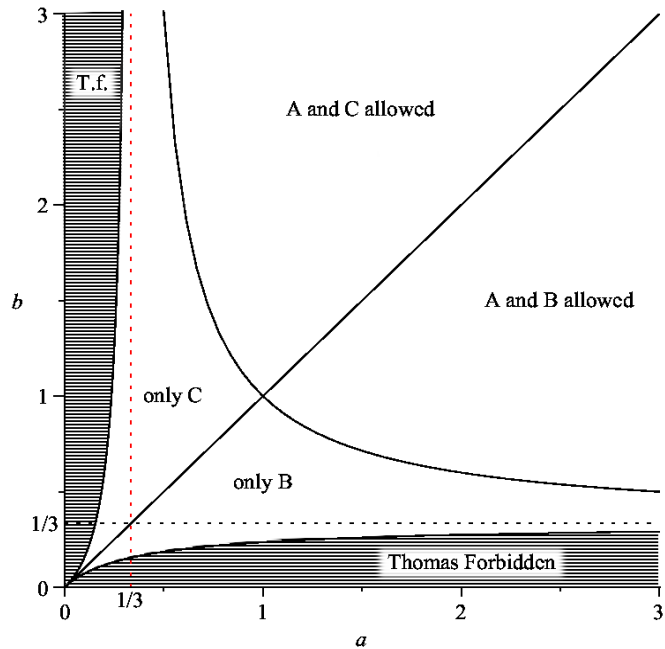}
\end{center}
\end{figure}

\newpage
\section{Capture processes with three or more collisions}
\subsection{Preliminaries}
In order for a collision sequence to lead to capture, a few simple
conditions must be met:
\\
1. Consecutive collisions must be distinct since two isolated particles
cannot collide more than once.\\
2. The first collision cannot be (1) because particle 1 initiates the
capture process and the last collision must not be (3) because if 1 and 2
collide at the end of the process they cannot emerge with the same velocity.\\
3. The times between collisions must be positive.

Using these conditions, we prove the following useful lemma in Appendix A.
\begin{lemma}
When three distinct collisions between three particles occur consecutively, a fourth collision must be of the same type as the second.  Furthermore, sequences of the form (31323) in which the first, third and fifth collisions are of the same type, but the second and fourth are of a different type, cannot occur.
\end{lemma}

Since capture begins with 2 and 3 having the same velocity and ends with 1
and 2 having the same velocity, the collision sequence can be considered to
begin with the collision (1) and to end with the collision (3). If three
distinct collisions were to occur consecutively between the virtual (1) and
virtual (3), our lemma would be violated. Therefore a capture process must
only consist of two types of collisions. There are only five such processes,
namely: $(32)^n,$ $(2)(32)^n,$ $(21)^n,$ $(21)^n (2),$ and $(31)^n.$ This is
fortuitous because the classical scattering matrices for such processes are
computed simply by raising to a power the scattering matrix of the repeated
pair.

We will first consider $(32)^n$ and $(2)(32)^n,$ which have allowed regions
which are within the forbidden region of the Lieber diagram. Because each
collision in these processes involves particle 1, the whole problem can be
done in the rest frame of 1. The scattering matrices are easily found for
such a frame, but since particle 1 accelerates when it collides
in an inertial frame and in the frame we are describing, it is at rest, this frame is noninertial.  Thus energy and momentum will not be conserved in collisions (2) and (3) (the ones in which particle 1 is involved). In this
scheme, one simply applies the scattering matrices to the velocity vector
and then subtracts from each of the three velocities, the velocity of
particle 1. If we have a matrix $M$ and an initial velocity vector $\vec{v_i}%
,$ the final velocity in the particle 1 frame, $\vec{v_f}$ is given by:

\begin{eqnarray}
\vec{v_f}&=&M\vec{v_i}- \left(
\begin{array}{c}
1 \\
1 \\
1
\end{array}
\right) \left[ \left(
\begin{array}{ccc}
1 & 0 & 0
\end{array}
\right) M\vec{v_i} \right] \\
&=& \left[I-\left(
\begin{array}{c}
1 \\
1 \\
1
\end{array}
\right) \left(
\begin{array}{ccc}
1 & 0 & 0
\end{array}
\right) \right]M\vec{v_i}
\end{eqnarray}

So to get the correct matrix one multiplies the matrix $M$ by
\[
\left(
\begin{array}{ccc}
0 & 0 & 0 \\
-1 & 1 & 0 \\
-1 & 0 & 1
\end{array}
\right).
\]
The benefit of working in the frame of particle 1 is that only two
velocities need be dealt with and consequently the scattering matrices
become two by two matrices, the first row and first column being irrelevant.
The scattering matrices, $S2$ and $S3$, are in this representation given by:

\begin{equation}
S2=\left(
\begin{array}{cc}
1 & b(y-1)/(a+b) \\
0 & y
\end{array}
\right)\mbox{, }S3=\left(
\begin{array}{cc}
x & 0 \\
(x-1)/(a+1) & 1
\end{array}
\right)
\end{equation}

One needs to know what the scattering angle is at each step to find the $x$s
and $y$s. To that end, the following lemma, proved in Appendix B, is useful.
\begin{lemma}
In a process of the form $(2)(32)^n$ or $(32)^n$, all scattering angles, excluding the first, are equal.
\end{lemma}

\newpage
\subsection{Collisions of type $(32)^n$}
The first step in analyzing $(32)^n$ or $(2)(32)^n$ is to find the relation between the first and second scattering angles.  Let the first and second scattering phasors be
$x=\exp(i\alpha)$ and $y=\exp(i\beta).$ Suppose the particles 2 and 3 initially have unit velocity.  Then following the first collision, the velocities of 2 and 3 are $x$ and $(a+x)/(a+1).$ The second collision is between 3 and 1 and we should like to know its scattering angle, $\phi.$ If we exponentiate the equation derived above for scattering angle, we get:
$y=\exp(i\beta)=-\exp(2i\psi)/\exp(2i\phi).$ In the present case,
 $\exp(2i\phi)=[(a+x)/(a+1)]/[(a+x)/(a+1)]^*=x(a+x)/(ax+1)$ and $\exp(i\psi)=x,$ so that:
\begin{equation}
y=-x(ax+1)/(a+x).  \label{yrelx}
\end{equation}
Now there is only one unknown variable in the problem, namely $x.$ It is
found by insisting that, in the particle-1 frame, 2 have zero velocity
finally. To find 2's final velocity, we need to multiply the string of
scattering matrices. Since only the first collision has a unique scattering
angle, the latter $2n-2$ collision matrices may be expressed as a power to
which the first two scattering matrices are appended:
\begin{equation}
S=(S_2 S_3)^{(n-1)}S_2^{(\beta/\pi)} S_3^{(\alpha/\pi)}=(S_2 S_3)^n
S_3^{(\alpha-\beta)/\pi}.
\end{equation}

To exponentiate $S_2 S_3,$ we need to diagonalize it.
\begin{equation}
S_2 S_3=\left(
\begin{array}{cc}
y+b(y-1)^2/(a+b)(a+1) & b(y-1)/(a+b) \\
y(y-1)/(a+1) & y
\end{array}
\right).
\end{equation}

The trace and determinant of the matrix are given by:
\begin{eqnarray}
\mbox{Trace}(S_2 S_3)&=&2y\left[1+\frac{b(y-1)^2}{2y(a+1)(a+b)}\right] \\
\det(S_2 S_3)&=&y^2.
\end{eqnarray}
Since the determinant is the product of the eigenvalues, the eigenvalues may
be expressed as $\epsilon_1=y/g$ and $\epsilon_2=y\cdot g,$ where $g$ is to
be determined. Furthermore, since the trace is the sum of these, we have:
\begin{equation}
2y\cdot\frac{1}{2}(g+1/g) =2y\left[1+\frac{b(y-1)^2}{2y(a+1)(a+b)}\right].
\end{equation}
This equation suggests that we write $g=e^{i\theta}$ with $\theta$ defined
by:
\begin{equation}
\cos(\theta)=1+\frac{b(y-1)^2}{2y(a+1)(a+b)} .  \label{defnoftheta}
\end{equation}
It will now be shown that $\theta$ is real and that the substitution is
justified. From the identity $(y-1)^2/y=-4\sin^2(\beta/2),$ we can write $%
\cos(\theta)=1-2b\sin^2(\beta/2)/((a+1)(a+b)),$ so that:

\begin{eqnarray*}
(a+1)(a+b)\ge b&\Rightarrow& \\
1\ge \frac{b}{(a+1)(a+b)}\ge \frac{b\sin^2(\beta/2)}{(a+1)(a+b)}
&\Rightarrow& \\
-1\le 1-\frac{2b\sin^2(\beta/2)} {(a+1)(a+b)}\le 1 &\Rightarrow& \\
-1\le\cos(\theta)\le 1.&&
\end{eqnarray*}
So $\theta$ is real and the eigenvalues may be expressed as $%
\epsilon_\pm=y\exp(\pm i\theta)$.  S may be decomposed into idempotent
matrices, $P_+$ and $P_-$ as $S=\epsilon_+ P_+ +\epsilon_- P_-,$ if $P_+
=(S-\epsilon_-I)/(\epsilon_+ -\epsilon_-)$ and $P_-
=(S-\epsilon_+I)/(\epsilon_- -\epsilon_+).$

We are now in a position to state the condition for capture. The final
velocity of particle 2 is $0,$ so the condition is:

\begin{equation}
\left(
\begin{array}{cc}
1 & 0
\end{array}
\right)(S_2S_3)^n S_3^{(\alpha-\beta)/\pi} \left(
\begin{array}{c}
1 \\
1
\end{array}
\right)=0.  \label{condition}
\end{equation}

Inserting the eigenvalue expansion, we have:

\begin{eqnarray}
0&=&\left(
\begin{array}{cc}
1 & 0
\end{array}
\right)\left[\epsilon_+^n\frac{S-\epsilon_-I}{\epsilon_+-\epsilon_-}%
+\epsilon_-^n\frac{S-\epsilon_+I}{\epsilon_--\epsilon_+}\right]%
S_3^{(\alpha-\beta)/\pi} \left(
\begin{array}{c}
1 \\
1
\end{array}
\right)  \label{sincond} \\
&=&\left(
\begin{array}{cc}
1 & 0
\end{array}
\right) \frac{y^{n-1}}{e^{i\theta}-e^{-i\theta}} \left[e^{in\theta}(S-ye^{-i%
\theta}I)- e^{-in\theta}(S-ye^{i\theta}I)\right] S_3^{(\alpha-\beta)/\pi}
\left(
\begin{array}{c}
1 \\
1
\end{array}
\right) \\
&=& \left(
\begin{array}{cc}
1 & 0
\end{array}
\right)\frac{y^{n-1}}{\sin(\theta)} \left[\sin(n\theta)S_2^{(\beta/%
\pi)}S_3^{(\beta/\pi)}-y\sin((n-1)\theta)I\right]S_3^{(\alpha-\beta)/\pi}
\left(
\begin{array}{c}
1 \\
1
\end{array}
\right)  \label{sineq}
\end{eqnarray}

When the appropriate matrix elements of $S_2^{(\beta/\pi)}S_3^{(\beta/\pi)}$
and $S_3^{(\alpha-\beta)/\pi}$ are inserted into Eq.~(\ref{sineq}), we
obtain:
\begin{equation}
\frac{\sin((n-1)\theta)}{\sin(n\theta)}=1-\frac{2b(a\cos\alpha+1)}{(a+b)(a+1)%
}.  \label{thetalpha}
\end{equation}

It will be useful to eliminate the alpha dependence from this equation.
First we invert Eq.~(\ref{yrelx}). As it is quadratic in $x,$ there are two
solutions:

\begin{eqnarray}
x&=&\left[-(1+y)\pm\sqrt{(1+y)^2-4a^2y}\right]/(2a)\Rightarrow  \label{xrely}
\\
\cos(\alpha)&=&\left[-\cos^2(\beta/2)\pm\sin(\beta/2)\sqrt{%
a^2-\cos^2(\beta/2)}\right]/a.
\end{eqnarray}
If either solution is inserted into Eq.~(\ref{thetalpha}), and Eq.~(\ref
{defnoftheta}) is applied, we obtain the following relation:
\begin{equation}
\frac{\cos^2(\theta/2)}{\sin^2(n\theta)}=\frac{a(1+b)}{(a+b)}.  \label{alpha}
\end{equation}

Now for given $a$ and $b,$ there may be several solutions to
Eq.~(\ref{alpha}).  We can determine their legitimacy by looking at the signs of the time intervals between collisions.  We construct a formula for the ratios of successive times.  Consider the initial positions
and velocities of the particles. If the process occurs in the x-y plane,
with 2 at the origin and 1's velocity in the positive x direction, and if
the polar angle of 3's position vector is $\phi$ and the distance between 2
and 3 is unity, then the initial position vector is $\left(
\begin{array}{cc}
0 & e^{i\phi}
\end{array}
\right).$ The position vector at the moment of the $(2j)^{th}$ collision is:
\begin{eqnarray}
\mathbf{r}_{2j}&=&\left(
\begin{array}{c}
0 \\
e^{i\phi}
\end{array}
\right)+ t_1S_3^{\alpha/\pi}\left(
\begin{array}{c}
1 \\
1
\end{array}
\right)+t_2S_2^{\beta/\pi}S_3^{\alpha/\pi}\left(
\begin{array}{c}
1 \\
1
\end{array}
\right)+  \label{even} \\
&&\cdots
t_{2j-1}\left(S_3^{\beta/\pi}S_2^{\beta/\pi}\right)^{j-1}S_3^{\alpha/\pi}%
\left(
\begin{array}{c}
1 \\
1
\end{array}
\right).  \nonumber
\end{eqnarray}

At the moment of the $(2j+1)^{th}$ it is:
\begin{eqnarray}
\mathbf{r}_{2j+1}&=&\left(
\begin{array}{c}
0 \\
e^{i\phi}
\end{array}
\right)+ t_1S_3^{\alpha/\pi}\left(
\begin{array}{c}
1 \\
1
\end{array}
\right)+ t_2S_2^{\beta/\pi}S_3^{\alpha/\pi}\left(
\begin{array}{c}
1 \\
1
\end{array}
\right)+  \label{odd} \\
&&\cdots t_{2j}\left(S_2^{\beta/\pi}S_3^{\beta/\pi}\right)^j
S_3^{(\alpha-\beta)/\pi}\left(
\begin{array}{c}
1 \\
1
\end{array}
\right).  \nonumber
\end{eqnarray}

The position vectors of relation (\ref{even}) must have their second
component zero because particle 3 is at the origin at the moment of an even
numbered collision. Similarly, the position vectors of relation (\ref{odd})
must have their first component zero because particle two is at the origin
at the moment of an odd numbered collision. From these requirements, we can
write equations from which to determine the times between collisions. Their
basic form is:
\begin{equation}
\left(
\begin{array}{cc}
0 & 1
\end{array}
\right)\mathbf{r}_{2j}=0\mbox{ and } \left(
\begin{array}{cc}
1 & 0
\end{array}
\right)\mathbf{r}_{2j+1}=0.
\end{equation}
First we will find the ratio between an even numbered time and the preceding
odd numbered time. If we subtract the equation $\left(
\begin{array}{cc}
1 & 0
\end{array}
\right)\mathbf{r}_{2j-1}=0$ from the equation $\left(
\begin{array}{cc}
1 & 0
\end{array}
\right)\mathbf{r}_{2j+1}=0,$ applying Eq.~(\ref{odd}), we find that:
\begin{eqnarray}
0&=&t_{2j-1}\left(
\begin{array}{cc}
1 & 0
\end{array}
\right)S_3^{\beta/\pi}\left(S_2^{\beta/\pi}S_3^{\beta/\pi}%
\right)^{j-1}S_3^{(\alpha-\beta)/\pi}\left(
\begin{array}{c}
1 \\
1
\end{array}
\right) \\
&&+t_{2j}\left(
\begin{array}{cc}
1 & 0
\end{array}
\right)\left(S_2^{\beta/\pi}S_3^{\beta/\pi}\right)^j
S_3^{(\alpha-\beta)/\pi}\left(
\begin{array}{c}
1 \\
1
\end{array}
\right).  \nonumber
\end{eqnarray}
If we apply the identity $\left(
\begin{array}{cc}
1 & 0
\end{array}
\right)S_3^{\beta/\pi}\mathbf{v}=y\left(
\begin{array}{cc}
1 & 0
\end{array}
\right)\mathbf{v},$ which holds for any velocity vector $\mathbf{v},$ and
carry out the exponentiations, we find:
\begin{equation}
\frac{t_{2j}}{t_{2j-1}}=\frac{\left(
\begin{array}{cc}
1 & 0
\end{array}
\right)\left[\sin((j-1)\theta)S_2^{(\beta/\pi)}
S_3^{(\alpha/\pi)}-y\sin((j-2)\theta)S_3^{(\alpha-\beta)/\pi}\right]\left(
\begin{array}{c}
1 \\
1
\end{array}
\right)}{\left(
\begin{array}{cc}
1 & 0
\end{array}
\right) \left[\sin((j)\theta)S_2^{(\beta/\pi)}
S_3^{(\alpha/\pi)}-y\sin((j-1)\theta)S_3^{(\alpha-\beta)/\pi}\right]\left(
\begin{array}{c}
1 \\
1
\end{array}
\right)}.  \label{trato}
\end{equation}
The ratio of the two matrix elements of the numerator and denominator can be
extracted from Eq.~(\ref{sincond}):
\begin{equation}
\frac{\left(
\begin{array}{cc}
1 & 0
\end{array}
\right)S_2^{(\beta/\pi)}S_3^{(\alpha/\pi)} \left(
\begin{array}{c}
1 \\
1
\end{array}
\right)} {\left(
\begin{array}{cc}
1 & 0
\end{array}
\right)S_3^{(\alpha-\beta)/\pi}\left(
\begin{array}{c}
1 \\
1
\end{array}
\right)}=y\frac{\sin((n-1)\theta)}{\sin(n\theta)}.
\end{equation}
If this ratio is applied to Eq.~(\ref{trato}), it can be shown that:
\begin{equation}
\frac{t_{2j}}{t_{2j-1}}=-\frac{\sin((n-j+1)\theta)}{\sin((n-j)\theta)},\hspace{1 cm}    j=1,2, \cdots (n-1)
\label{sinrat}
\end{equation}

It remains to find the ratio of an odd numbered time to the preceding even
numbered time. We subtract $\left(
\begin{array}{cc}
0 & 1
\end{array}
\right)\mathbf{r}_{2j}=0$ from $\left(
\begin{array}{cc}
0 & 1
\end{array}
\right)\mathbf{r}_{2j+2}=0,$ to find:
\begin{eqnarray}
0&=&t_{2j}\left(
\begin{array}{cc}
0 & 1
\end{array}
\right)S_2^{\beta/\pi}\left(S_3^{\beta/\pi}S_2^{\beta/\pi}\right)^j
S_3^{\alpha/\pi}\left(
\begin{array}{c}
1 \\
1
\end{array}
\right) \\
&&+t_{2j+1}\left(
\begin{array}{cc}
0 & 1
\end{array}
\right)\left(S_3^{\beta/\pi}S_2^{\beta/\pi}\right)^{j+1}
S_3^{\alpha/\pi}\left(
\begin{array}{c}
1 \\
1
\end{array}
\right).  \nonumber
\end{eqnarray}
Following the same procedure as above, we use the fact that $\left(
\begin{array}{cc}
0 & 1
\end{array}
\right)$ is a left eigenvector of $S_2^{\beta/\pi}$ and that $S_3S_2$ and $%
S_2S_3$ have the same eigenvalues to derive:
\begin{equation}
\frac{t_{2j+1}}{t_{2j}}=-\frac{\left(
\begin{array}{cc}
0 & 1
\end{array}
\right)\left[\sin((j-1)\theta)S_3^{\beta/\pi}S_2^{\beta/\pi}S_3^{\alpha/%
\pi}-y\sin((j-2)\theta)S_3^{\alpha/\pi}\right]\left(
\begin{array}{c}
1 \\
1
\end{array}
\right)}{\left(
\begin{array}{cc}
0 & 1
\end{array}
\right)\left[\sin(j\theta)S_3^{\beta/\pi}S_2^{\beta/\pi}S_3^{\alpha/\pi}-y%
\sin((j-1)\theta)S_3^{\alpha/\pi}\right]\left(
\begin{array}{c}
1 \\
1
\end{array}
\right)} .  \label{trate}
\end{equation}
As before, we first evaluate the ratio of the matrix elements, $\rho.$ If we
perform the matrix multiplications and apply Eqs.~(\ref{yrelx}) and (\ref
{defnoftheta}), we find that:
\begin{equation}
\rho=\frac{\left(
\begin{array}{cc}
0 & 1
\end{array}
\right)S_3^{(\beta/\pi)}S_2^{(\beta/\pi)}S_3^{(\alpha/\pi)} \left(
\begin{array}{c}
1 \\
1
\end{array}
\right)} {\left(
\begin{array}{cc}
0 & 1
\end{array}
\right)S_3^{(\alpha/\pi)}\left(
\begin{array}{c}
1 \\
1
\end{array}
\right)}=y\left[2\cos\theta-1+\frac{2(a\cos\alpha+1)}{a^2+2a\cos\alpha+1}%
\right].
\end{equation}
Next we solve Eq.~(\ref{thetalpha}) for $2(a\cos\alpha+1)$ in terms of $%
\theta,$ $a,$ and $b,$ and use the result to simplify $\rho:$
\begin{equation}
\rho=y\left[2\cos\theta-1+\frac{\sin(n\theta)-\sin((n-1)\theta)}{%
[a(b+1)/(a+b)]\sin(n\theta)-\sin((n-1)\theta)}\right].
\end{equation}
Finally, if we eliminate $a(b+1)/(a+b)$ using Eq.~(\ref{alpha}), we obtain:
\begin{equation}
\rho=y\cos((n-3/2)\theta)/\cos((n-1/2)\theta).
\end{equation}
And when the expression for $\rho$ is used in Eq.~(\ref{trate}), it can be
shown that:
\begin{equation}
\frac{t_{2j+1}}{t_{2j}}=-\frac{\cos((n-j+1/2)\theta)}{\cos((n-j-1/2)\theta)},\hspace{1 cm}    j=1,2, \cdots (n-1).
\label{cosrat}
\end{equation}
As was mentioned earlier, for a given $a$ and $b,$ there may be several
values of $\theta$ which satisfy Eq.~(\ref{alpha}) but at most one
corresponds to a process with positive times between collisions. The time
between the first two collisions, $t_1$ can be guaranteed to be positive
because after 1 collides with 2, a collision with 3 is possible if 3 has the
right orientation. But from relations (\ref{sinrat}) and (\ref{cosrat}), we see
that for all the times to be positive, we need $\sin((n-j+1)\theta)/
\sin((n-j)\theta)<0$ and $\cos((n-j+1/2)\theta)/
\cos((n-j-1/2)\theta)<0$ for all $1\le j\le n-1.$

The solution set for $\sin(k\theta)/\sin((k-1)\theta)<0$ is
\begin{equation}
\bigcup_{l=1}^{k-1} (l\pi/k,l\pi/(k-1))\label{firstset}
\end{equation}
and the solution set for $\cos((k+1/2)\theta)/
\cos((k-1/2)\theta)<0$ is
\begin{equation}
\bigcup_{l=1}^{k} ((2l-1)\pi/(2k+1),(2l-1)\pi/(2k-1)).\label{secondset}
\end{equation}
In (\ref{firstset}) and (\ref{secondset}), $k=n-j$ and $j$ varies from $1$ to $n-1$ so $k$ can take any value between 1 and $n-1$.  It is not hard to show that the intersection of the sets (\ref{firstset}) having $k$ between $1$ and $n-1$ is $((n-2)\pi/(n-1),\pi)$.  The intersection of this interval with all the sets (\ref{secondset}) having $k$ between $1$ and $n-1$ is $((2n-3)\pi/(2n-1),\pi)$.  Thus a necessary condition for all positive time intervals is that $\theta$ be in the range $((2n-3)\pi/(2n-1),\pi)$.

We have seen that $\theta$ is real if $\beta$ is. From the form of Eq.~(\ref{yrelx}), the reality of $\alpha$ guarantees
the reality of $\beta.$

All the dynamical variables of the problem are found
by multiplying the original velocity vector by the $2\times 2$ matrices $S_2$
and $S_3.$ So a process is classically allowed if $\alpha\in\Re$ and $(2n-3)\pi/(2n-1)\le\theta\le\pi.$ Eq.~(\ref{xrely}) may be rewritten as $x=\sqrt{y}\exp(i\phi),$ if $\cos\phi=-(1+y)/(2a\sqrt{y})=-\cos(\beta/2)/a.$ In this
form, it is evident that if $\beta\in\Re,$ then $\alpha\in\Re,$ iff $\cos^2(\beta/2)\le a^2.$ This inequality may be recast in terms of sines as:
\begin{equation}
1-a^2\le\sin^2(\beta/2)\le 1\Leftrightarrow \frac{b(1-a)}{(a+b)}
\le\sin^2(\theta/2)\le\frac{b}{(a+b)(a+1)}.
\end{equation}
The lower bound on $\sin^2(\theta/2)$ poses no constraint, it being guaranteed by Eq.~(\ref{alpha}), so we must solve that equation in the domain
$(2n-3)\pi/(2n-1)\le\theta\le\pi,$ subject to the constraint
\begin{equation}
\sin^2(\theta/2)\le b/((a+b)(a+1)).\label{mainineq}
\end{equation}
Let $Q(\theta)$ be defined by $Q(\theta)=\cos^2(\theta/2)/\sin^2(n\theta)$.  Then Eq.~(\ref{alpha}) can be written as
\begin{equation}
\theta=Q^{-1}\left(\frac{a(1+b)}{a+b}\right).
\end{equation}
This equation can be used to assign a value to $\theta$ at every point in the $ab$ plane at which $Q^{-1}$ exists.  On the interval $((2n-3)\pi/(2n-1),\pi)$, $Q(\theta)$ decreases monotonically from $\infty$ to $1/4n^2$.  Thus, we may only choose $a$ and $b$ for which
\begin{equation}
\frac{a(1+b)}{a+b}>1/4n^2.\label{minorineq}
\end{equation}
The points $(a, b)$ which satisfy the inequality $a(1+b)/(a+b)>Q_0$ lie between the branches of a hyperbola which passes through the origin and has asymptotes at $a=1/Q_0$ and $b=1/Q_0-1$.  Because $Q$ decreases monotonically on the interval of interest, if $\theta_1$, $\theta_2$ and $\theta_3$ satisfy $(2n-3)\pi/(2n-1)\le\theta_1<\theta_2<\theta_3\le\pi$, it follows that if $\theta_1\le\theta(a,b_1)\le\theta_2$ and $\theta_2\le\theta(a,b_2)\le\theta_3$, then $b_2\ge b_1$.  In other words, the region in which $\theta_1\le\theta\le\theta_2$ lies under the region in which $\theta_2\le\theta\le\theta_3$.

We can find the allowed region for this process by looking for the curves along which the left and right hand sides of the inequality (\ref{mainineq}) are equal.  Then we can test to see on which side of the curves the inequality is met.  Finally we must check the inequality (\ref{minorineq}) to make sure that $\theta$ actually exists in the proposed region.

We write \ref{mainineq} as an equation:
\begin{equation}
\sin^2(\theta/2)= b/((a+b)(a+1)).\label{maineq}
\end{equation}
and combine it with Eq.~(\ref{alpha}) to obtain:
\begin{equation}
\tan^2(n\theta)= \frac{1+a+b}{ab}.\label{ntan}
\end{equation}

Also, Eq.~(\ref{maineq}) can be modified to yield:
\begin{equation}
\tan^2(\theta/2)=\frac{b}{a(1+a+b)}\label{tanover2}.
\end{equation}

If we multiply Eqs.~(\ref{ntan}) and (\ref{tanover2}) together and solve for $a$, we get two solutions.  The two solutions for $a$ along with the corresponding expressions for $b$ are as follows:
\begin{eqnarray}
a&=&\cot(\theta/2)\cot(n\theta) \label{a1}\\
b&=&\cot(n\theta)\cot((n-1/2)\theta)\mbox{, if} \\
&&(2n-2)\pi/(2n-1)\le\theta\le(2n-1)\pi/(2n)  \nonumber
\end{eqnarray}
and
\begin{eqnarray}
a&=&-\cot(\theta/2)\cot(n\theta) \label{a2}\\
b&=&\cot(n\theta)\cot((n+1/2)\theta)\mbox{, if} \\
&&(2n-1)\pi/(2n)\le\theta\le(2n)\pi/(2n+1).  \nonumber
\end{eqnarray}
The limits on $\theta$ are determined by finding the values of $\theta$ which make $a$ and $b$ both positive.
For convenience, let the former be denoted by $C_{2n-1}$ and the latter by $C_{2n}$.
If we let $\theta_1=(2n-2)\pi/(2n-1)$, $\theta_2=(2n-1)\pi/2n$ and $\theta_3=2n\pi/(2n+1)$, then it follows from the above discussion that $C_{2n-1}$ lies beneath $C_{2n}$.
By plugging $\theta=\theta_1$ into (\ref{a1}), we find that $C_{2n-1}$ has an asymptote at $a=\tan^2(\pi/(4n-2))$ and by plugging $\theta=\theta_3$ into (\ref{a2}), we find that $C_{2n}$ has an asymptote at $a=\tan^2(\pi/(4n+2))$.
To test whether the process is allowed between these two or outside them, we study a test point.  If $b=0$ and $a>0$, then $\theta$ exists and is given by $Q^{-1}(1)$.  Since $(2n-3)\pi/(2n-1)\le\theta\le\pi$, it follows that $\sin^2(\theta/2)\ge \cos^2(\pi/2n)$ but from inequality (\ref{mainineq}) we need $\sin^2(\theta/2)\le 0$ at such a point.  Thus the allowed region for the process is between the curves $C_{2n-1}$ and $C_{2n}$.

\newpage
\subsection{Collisions of type $(2)(32)^n$}
The next process to consider is $(2)(32)^n.$ The relation between the first and
following scattering angles is different for that problem than for $(32)^n.$
In the above derivation of the relation, only the effects of the first two
collisions are considered so the equation for the $(2)(32)^n$ problem is
obtained by exchanging the masses of 2 and 3 in Eq.~(\ref{yrelx}):
\begin{equation}
y=-x(ax+b)/(a+bx).  \label{yrelx2}
\end{equation}
The condition for capture is:

\begin{eqnarray}
0&=&\left(
\begin{array}{cc}
1 & 0
\end{array}
\right)(S_2S_3)^n S_2^{\alpha/\pi} \left(
\begin{array}{c}
1 \\
1
\end{array}
\right)  \label{condition232} \\
&=&\left(
\begin{array}{cc}
1 & 0
\end{array}
\right)\left[\sin(n\theta)S_2^{\beta/\pi}S_3^{\beta/\pi}S_2^{\alpha/\pi}-y%
\sin((n-1)\theta)S_2^{\alpha/\pi}\right]\left(
\begin{array}{c}
1 \\
1
\end{array}
\right).
\end{eqnarray}

If we evaluate the matrix elements with the aid of Eq.(~\ref{defnoftheta}),
we find that:
\begin{equation}
\frac{\sin(n\theta)}{\sin((n+1)\theta)}=\frac{a^2+2ab\cos\alpha+b^2}{a^2-b^2}.  \label{hasalpha2}
\end{equation}
A formula is needed for $\alpha$ given $\beta$. This is obtained by
exchanging the masses of 2 and 3 in Eq.~(\ref{xrely}):
\begin{equation}
\cos{\alpha}=\left[-b\cos^2(\beta/2)\pm\sqrt{a^2-b^2\cos^2(\beta/2)}%
\sin(\beta/2)\right]/a.
\end{equation}
This formula, along with Eq.~(\ref{defnoftheta}) allows us to rewrite Eq.~(%
\ref{hasalpha2}) as:
\begin{equation}
\frac{\cos^2((n+1/2)\theta)}{\cos^2(\theta/2)}=\frac{b(1+a)}{a(1+b)}
\label{alpha2}
\end{equation}

As with $(32)^n$, we next determine which values of $\theta$ can be ruled
out because of negative time intervals between collisions. Since 3 begins at
the origin, $\phi$ will now denote the initial polar angle of 2. The
position vector at the moment of the $(2j)^{th}$ collision is:
\begin{eqnarray}
\mathbf{r}_{2j}&=&\left(
\begin{array}{c}
e^{i\phi} \\
0
\end{array}
\right)+ t_1S_2^{\alpha/\pi}\left(
\begin{array}{c}
1 \\
1
\end{array}
\right)+ t_2S_3^{\beta/\pi}S_2^{\alpha/\pi}\left(
\begin{array}{c}
1 \\
1
\end{array}
\right)+  \label{evenprime} \\
&&\cdots
t_{2j-1}\left(S_2^{\beta/\pi}S_3^{\beta/\pi}\right)^{j-1}S_2^{\alpha/\pi}%
\left(
\begin{array}{c}
1 \\
1
\end{array}
\right).  \nonumber
\end{eqnarray}

At the moment of the $(2j+1)^{th}$ it is:
\begin{eqnarray}
\mathbf{r}_{2j+1}&=&\left(
\begin{array}{c}
e^{i\phi} \\
0
\end{array}
\right)+ t_1S_2^{\alpha/\pi}\left(
\begin{array}{c}
1 \\
1
\end{array}
\right)+ t_2S_3^{\beta/\pi}S_2^{\alpha/\pi}\left(
\begin{array}{c}
1 \\
1
\end{array}
\right)+  \label{oddprime} \\
&&\cdots t_{2j}\left(S_3^{\beta/\pi}S_2^{\beta/\pi}\right)^j
S_2^{(\alpha-\beta)/\pi}\left(
\begin{array}{c}
1 \\
1
\end{array}
\right).  \nonumber
\end{eqnarray}
As before, we derive the time ratios from the constraint that the two
particles involved in a collision have the same position. The equations
expressing that constraint are:
\begin{equation}
\left(
\begin{array}{cc}
1 & 0
\end{array}
\right)\mathbf{r}_{2j}=0\mbox{ and } \left(
\begin{array}{cc}
0 & 1
\end{array}
\right)\mathbf{r}_{2j+1}=0.
\end{equation}
Now we can use Eq.~(\ref{oddprime}) to find the ratio of an even numbered
time to the preceding odd numbered one:

\begin{equation}
\frac{t_{2j}}{t_{2j-1}}=-\frac{\left(
\begin{array}{cc}
0 & 1
\end{array}
\right) \left[\sin((j-1)\theta)S_3^{\beta/\pi}S_2^{\alpha/\pi}
-y\sin((j-2)\theta)S_2^{(\alpha-\beta)/\pi}\right]\left(
\begin{array}{c}
1 \\
1
\end{array}
\right) }{\left(
\begin{array}{cc}
0 & 1
\end{array}
\right)\left[\sin(j\theta)S_3^{\beta/\pi}S_2^{\alpha/\pi}
-y\sin((j-1)\theta)S_2^{(\alpha-\beta)/\pi}\right]\left(
\begin{array}{c}
1 \\
1
\end{array}
\right)}.  \label{232even}
\end{equation}
The relevant matrix element ratio is:
\begin{equation}
\rho=\frac{\left(
\begin{array}{cc}
0 & 1
\end{array}
\right) S_3^{\beta/\pi}S_2^{\alpha/\pi}\left(
\begin{array}{c}
1 \\
1
\end{array}
\right)}{\left(
\begin{array}{cc}
0 & 1
\end{array}
\right)S_2^{(\alpha-\beta)/\pi}\left(
\begin{array}{c}
1 \\
1
\end{array}
\right)}=y\left[1-\frac{2(b+a\cos\alpha)}{(a+1)(a+b)}\right].
\end{equation}

With the help of Eqs.~(\ref{hasalpha2}) and (\ref{alpha2}), we may eliminate
$a,$ $b,$ and $\alpha$ from this expression: $\rho=y\cos((n-1/2)\theta)/%
\cos((n+1/2)\theta).$ When this is used in Eq.~(\ref{232even}), we obtain:
\begin{equation}
\frac{t_{2j}}{t_{2j-1}}=-\frac{\cos((n-j+3/2)\theta)}{\cos((n-j+1/2)\theta)},\hspace{1 cm}    j=1,2, \cdots n.
\label{cosrat2}
\end{equation}
From Eq.~(\ref{evenprime}), the ratio of an odd numbered time to the
preceding even one is given by:
\begin{equation}
\frac{t_{2j+1}}{t_{2j}}=-\frac{y\left(
\begin{array}{cc}
1 & 0
\end{array}
\right) \left(S_2^{\beta/\pi}S_3^{\beta/\pi}
\right)^{j-1}S_2^{\alpha/\pi}\left(
\begin{array}{c}
1 \\
1
\end{array}
\right)}{\left(
\begin{array}{cc}
1 & 0
\end{array}
\right) \left(S_2^{\beta/\pi}S_3^{\beta/\pi} \right)^j S_2^{\alpha/\pi}\left(
\begin{array}{c}
1 \\
1
\end{array}
\right)}.
\end{equation}
Invoking Eq.~(\ref{condition232}), we can remove the $a,$ $b,$ and $\alpha$
dependence:
\begin{equation}
\frac{t_{2j+1}}{t_{2j}}=-\frac{\sin((n-j+1)\theta)}{\sin((n-j)\theta)},\hspace{1 cm}    j=1,2, \cdots (n-1).
\label{sinrat2}
\end{equation}
The solution set of $\cos((k+3/2)\theta)/\cos((k+1/2)\theta)<-1$ is
\begin{equation}
\bigcup_{l=1}^{k+1} ((2l-1)\pi/(2k+3),(2l-1)\pi/(2k+1)).\label{set1}
\end{equation}
and the solution set of $\sin((k+1)\theta)/\sin(k\theta)<-1$ is
\begin{equation}
\bigcup_{l=1}^k (l\pi/(k+1),l\pi/k)\label{set2}.
\end{equation}

In (\ref{set2}), $k=n-j$ and $j$ varies from $1$ to $n-1$ so $k$ can take any value between 1 and $n-1$.  It is not hard to show that the intersection of the sets (\ref{set2}) having $k$ between $1$ and $n-1$ is $((n-1)\pi/n,\pi)$.  In (\ref{set1}), $j$ varies from $1$ to $n$ so $k$ takes values between $0$ and $n-1$.  The intersection of $((n-1)\pi/n,\pi)$ with all the sets (\ref{set1}) having $k$ between $0$ and $n-1$ is $((2n-1)\pi/(2n+1),\pi)$.  Thus a necessary condition for all positive time intervals is that $\theta$ be in the range $((2n-1)\pi/(2n+1),\pi)$.

Eq.~(\ref{yrelx2}) may be inverted as
$x=\sqrt{y}\exp{i\phi},$ if
$\cos(\phi)=-b(1+y)/(2a\sqrt{y})=-b\cos(\beta/2)/a.$
We have $\phi\in\Re$ if $0\le\cos^2(\beta/2)\le a^2/b^2.$ Equivalently, we
may write the condition for capture as:
\begin{equation}
1-a^2/b^2\le\sin^2(\beta/2)\le 1\Leftrightarrow \frac{(b-a)}{b(a+1)}%
\le\sin^2(\theta/2)\le\frac{b}{(a+b)(a+1)}.  \label{Ineq2}
\end{equation}
The left inequality follows from Eq.~(\ref{alpha2}) so any solution to Eq.~(%
\ref{alpha2}) in the interval $(2n-1)\pi/(2n+1)\le\theta\le\pi$ satisfying
\begin{equation}
\sin^2(\theta/2)\le b/[(a+b)(a+1)]
\end{equation} corresponds to allowed capture. Let
$Q$ be defined by $Q(\theta)=\cos^2((n+1/2)\theta)/\cos^2(\theta/2).$ On the
interval under consideration, $Q$ increases monotonically in $\theta,$ from $%
0$ at the left endpoint to $(2n+1)^2$ at the right endpoint. From Eq.~(\ref
{alpha2}), we can show that $b=aQ/(1+a-aQ).$ When this is used in the
inequality we have:
\begin{equation}
\frac{Q}{(1+Q)+a(1-Q)}\ge (a+1)\sin^2(\theta/2).  \label{Qatheta}
\end{equation}
Taken as an equation, the two solutions of (\ref{Qatheta}) for $a$ are
$-\cot(\theta/2)\cot(n\theta)$ and $\cot(\theta/2)\cot((n+1)\theta).$
The corresponding expressions for $b$ and the values of $\theta$ which make $a$ and $b$ positive are as follows:
\begin{eqnarray}
a&=&-\cot(\theta/2)\cot(n\theta)\\
b&=&\cot((n+1/2)\theta)\cot(n\theta)\mbox{, if}\\
&&(2n-1)\pi/(2n) \le \theta \le 2n\pi/(2n+1)\nonumber
\end{eqnarray} and
\begin{eqnarray}
a&=&\cot(\theta/2)\cot((n+1)\theta)\\
b&=&\cot((n+1/2)\theta)\cot((n+1)\theta)\mbox{, if}\\
&&2n\pi/(2n+1) \le \theta \le (2n+1)\pi/(2n+2).\nonumber
\end{eqnarray}
The first curve is the same as $C_{2n}$ and the second is the same as $C_{2n+1}.$  Thus, the boundary curves
for $2(32)^n$ are the same as those for $(32)^n.$ The functions $a_1$ and $%
a_2$ are found by combining Eq.~(\ref{alpha2}) with the Eq.~(\ref{Ineq2}).
This has been done by eliminating $b$ from Eq.~(\ref{Ineq2}).

But a more useful form for the curves is obtained by eliminating $\theta.$ If Eq.~(\ref
{Ineq2}) is taken as an equality, then $\exp(i\theta/2)$ may be written as
\begin{equation}
\exp(i\theta/2)=\frac{\sqrt{a(a+1+b)}+i\sqrt{b}}{\sqrt{(a+b)(1+a)}}.
\label{thover2}
\end{equation}
From Eq.~(\ref{alpha2}), $\cos^2((n+1/2)\theta)=b(a+1+b)/((a+b)(1+b))$ and
\begin{equation}
\exp(i(n+1/2)\theta)=\frac{\sqrt{b(a+1+b)}\pm i\sqrt{a}}{\sqrt{(a+b)(1+b)}}.
\label{n+1/2}
\end{equation}
The sign of $\sin((n+1/2)\theta)$ depends on the parity of $n.$ For even $n,$
$\sin((n+1/2)\theta)$ is negative if $(2n-1)\pi/(2n)\le\theta\le 2n\pi/(2n+1)
$ and positive if $2n\pi/(2n+1)\le\theta\le (2n+1)\pi/(2n+2),$ while for odd
$n,$ the opposite is true. Combining Eqs.~(\ref{thover2}) and (\ref{n+1/2})
gives for $C_{2n}$:
\begin{equation}
\left(\frac{\sqrt{a(a+1+b)}+i\sqrt{b}}{\sqrt{a(a+1+b)}-i\sqrt{b}}%
\right)^{2n+1}=\frac{\sqrt{b(a+1+b)}-(-1)^n i\sqrt{a}}{\sqrt{b(a+1+b)}%
+(-1)^n i\sqrt{a}}\label{evencurve}
\end{equation}
and for $C_{2n+1}$:
\begin{equation}
\left(\frac{\sqrt{a(a+1+b)}+i\sqrt{b}}{\sqrt{a(a+1+b)}-i\sqrt{b}}%
\right)^{2n+1}=\frac{\sqrt{b(a+1+b)}+(-1)^n i\sqrt{a}}{\sqrt{b(a+1+b)}
-(-1)^n i\sqrt{a}}\label{oddcurve}
\end{equation}
The first four boundary curves are given by:
\begin{eqnarray}
0&=&b-a,\\
0&=&3ab-b+a^2+a,\\
0&=&3ab^2-b^2+2ba^2+6ab-a^3-a^2,\\
0&=&5a^2b^2-10ab^2+b^2+6ba^3-6ab+a^4+2a^3+a^2.
\end{eqnarray}
These curves are, respectively, linear, quadratic, cubic, and quartic.
Because the $(32)^n$ and $(2)(32)^n$ processes share boundary curves, the
allowed regions are contiguous and non-overlapping. Furthermore, the
asymptotes, having the basic form $a=\tan^2(\pi/(2(2m+1))),$ approach $a=0$
in the limit as $n$ approaches $\infty,$ and so the entire $a-b$ plane above
the line $b=a$ is covered by allowed regions for processes of these types.
The allowed regions for (32), (232), and (3232) are given in figure 3.
\begin{figure}[htp]
\caption{The pairs of curves which, going from right to left, give the allowed regions for (32), (232), (3232) and higher order processes.}
\begin{center}
\includegraphics[height=0.48\textheight,width=0.65\textwidth]{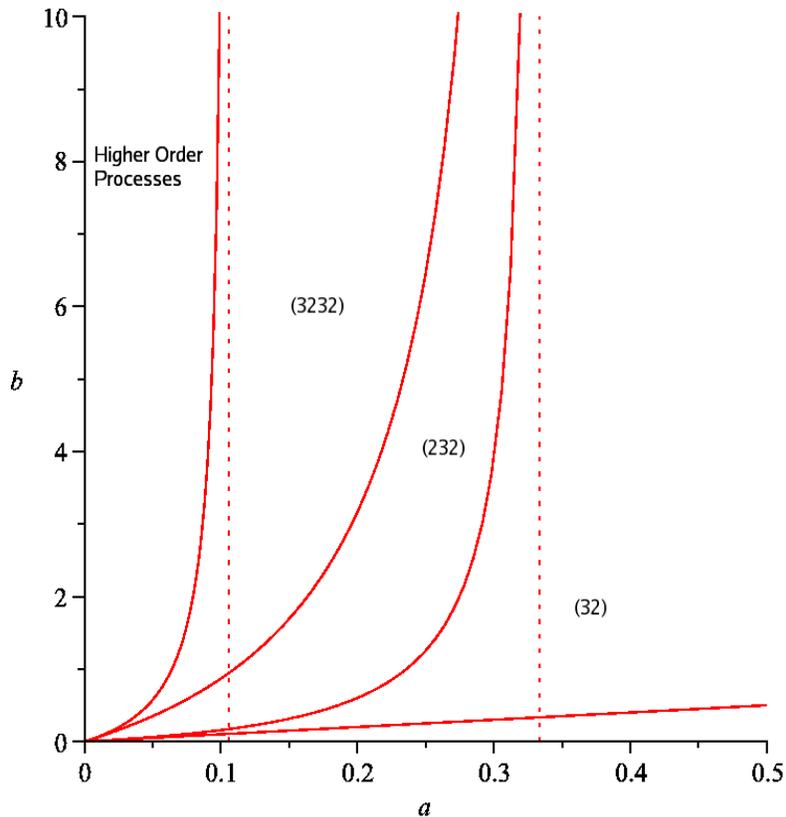}
\end{center}
\end{figure}
\\Starting at the a-axis and going counterclockwise, the figure shows curves $C_1$, $C_2$, $C_3$ and $C_4$.  In the figure we can see that the pair $C_1$ and $C_2$ as well as the pair $C_3$ and $C_4$ have the same 
slope at the origin.  Also $C_2$ and $C_3$ have the same asymptote.  It is not difficult to show from Eqs.~\ref{evencurve} and \ref{oddcurve} that, in general, $C_{2n+1}$ and $C_{2n+2}$ have the same slope at the 
origin and $C_{2n}$ and $C_{2n+1}$ share an asymptote.

A couple of examples of (2)(32)$^n$ type processes were mentioned in the introduction.  These and one other example are given in more detail now.  The capture of a proton from protonium by a muon is a (232) process:
\begin{equation}
\mu + (p\overline{p})\rightarrow(\mu p)+\overline{p}.
\end{equation}  The muon is particle 1, the proton is particle 2 and the antiproton is particle 3.  The scattering angle can be shown to be 2.72 rad.  The process Na + I$_2\rightarrow$ NaI + I is a molecular example of 
(232) which has scattering angle 1.09 rad.
The capture of a proton from muonic hydrogen by an electron:  $e + (\mu p)\rightarrow H+\mu$ is an example of a higher order process, (2)(32)$^{10}$, with scattering angle 1.09 rad (the same as the previous angle only to three significant figures).  The time reversed process, 
$H+\mu\rightarrow e + (\mu p)$ is more interesting because it is a process in which muonic hydrogen is created.  It is a (21)$^{10}$(2) process with scattering angle 59 mrad.

\newpage
\subsection{Collisions of type $(21)^n$ or $(21)^n(2)$}
We know that at the beginning of a capture process, 2 and 3 have zero
relative velocity and at the end, 1 and 2 have zero relative velocity. If we
examine a given process under time reversal, we see that, initially, 1 and 2
have zero relative velocity and, finally, 2 and 3 have zero velocity. If the
particle labels ``1" and ``3" are exchanged, one has a new capture process.
If this time reversal and label exchange is done to the process $(32)^n,$
what results is $(21)^n$ and when it is done to $(2)(32)^n,$ what results is $(21)^n(2).$ 
This means that if $(32)^n$ or $(2)(32)^n$ is allowed for a point $(a,b)=(a_0,b_0),$ then $(21)^n$ or $(21)^n(2),$ resp., is allowed for the
point $(a,b)=(b_0,a_0).$ Therefore the allowed regions for these processes
may be generated by reflecting the previously determined regions about the
line $b=a.$

\newpage
\subsection{Collisions of type $(31)^n$}
The $(31)^n$ problem is much simpler than the processes previously
considered because, as will be shown, each collision has the same scattering
angle. Because of this there will be only two variables to solve for as
opposed to the three in the above problem. Since 2 is involved in each collision,
we work in its frame. The scattering matrices are:

\begin{equation}
S_1=\left(
\begin{array}{cc}
1 & b(x-1)/(1+b) \\
0 & x
\end{array}
\right)\mbox{, } S_3=\left(
\begin{array}{cc}
y & 0 \\
a(y-1)/(1+a) & 1
\end{array}
\right)
\end{equation}

The condition which determines the relation between the first two scattering
angles is that the second collision (between 3 and 2) reverse particle 1's
velocity so that it can collide with 2. Or in the case of $(31)$, 1's velocity
is not reversed but made zero. In the frame of 2, the initial velocity is
$\left(
\begin{array}{cc}
1 & 0
\end{array}
\right)^T,$ the velocity following the first collision is: $\left(
\begin{array}{cc}
x & a(x-1)/(1+a)
\end{array}
\right)^T,$ and the final velocity is:

\begin{equation}
\left(
\begin{array}{c}
x+ab(x-1)(y-1)/(1+a)/(1+b) \\
ay(x-1)/(1+a)
\end{array}
\right).
\end{equation}

Before the second collision, 1 has velocity $x$ and following that
collision, it has velocity $x+ab(x-1)(y-1)/(1+a)/(1+b)$ so if 1's velocity
is to be reversed or brought to zero, $x$ and $(x-1)(y-1)$ must have
opposite phase. The phase of the former is $\alpha$ and that of the latter
is $\beta/2+\alpha/2+\pi,$ so the first and second scattering angles are
equal. This is demonstrated for the case $n=1$ by Bittensky, through a
geometrical argument \cite{ref2}. We can therefore find the total scattering
matrix simply by exponentiating $M=S_1^{\alpha/\pi} S_3^{\alpha/\pi}.$ M,
given by:

\begin{equation}
M=\left(
\begin{array}{cc}
x+ab(x-1)^2/(a+1)/(b+1) & b(x-1)/(1+b) \\
ax(x-1)/(1+a) & x
\end{array}
\right)
\end{equation}
has determinant $x^2$ and trace $2x+ab(x-1)^2/(a+1)/(b+1).$ As before, we
express the eigenvalues as $\epsilon_+ =xg$ and $\epsilon_- =x/g,$ with $%
g=\exp(i\theta)$ so that

\begin{equation}
\cos(\theta)=1+\frac{ab}{(a+1)(b+1)}\left(\frac{(x-1)^2}{2x}\right).
\label{defn31theta}
\end{equation}

The projection operators are:

\begin{eqnarray}
P_+ &=&\frac{M-\epsilon_-I}{\epsilon_+ -\epsilon_-} \\
P_- &=&\frac{M-\epsilon_+I}{\epsilon_- -\epsilon_+}.
\end{eqnarray}

Using these expressions and the above for $\epsilon_+$ and $\epsilon_-,$ we
find that
\begin{eqnarray}
M^n&=&\epsilon_+P_+ + \epsilon_-P_- \\
&=&\frac{x^{n-1}}{\sin(\theta)}(\sin(n\theta)M-x\sin((n-1)\theta)I)
\end{eqnarray}

The process is completed when 1 has zero velocity so the condition for
capture is:
\begin{eqnarray}
\left(
\begin{array}{cc}
1 & 0
\end{array}
\right)M^n\left(
\begin{array}{c}
1 \\
0
\end{array}
\right)=0 &\Rightarrow& \\
x\sin((n-1)\theta)/\sin(n\theta)&=&\left(
\begin{array}{cc}
1 & 0
\end{array}
\right)M\left(
\begin{array}{c}
1 \\
0
\end{array}
\right) \\
&=&x+ab(x-1)^2/(a+1)/(b+1).
\end{eqnarray}

When this equation is combined with the definition of $\cos(\theta),$ the
simple result: \newline
$\sin((n+1)\theta)=\sin(n\theta),$ emerges. Aside from the trivial solution $%
\theta=0,$ which corresponds to no collisions at all, there are n solutions:

\begin{equation}
\theta=(2m+1)\pi/(2n+1), m=0,1,2,\ldots,n  \label{thetaqu}
\end{equation}

The allowed region is now readily found by insisting that $\alpha$ be real,
which is to say, $-1 \le \cos(\alpha) \le 1.$ If we solve Eq.~(\ref
{defn31theta}) for $\cos(\alpha)$ and insist that it be real and between -1
and 1, we are left with:

\begin{equation}
1+a+b\le ab\cot^2(\theta/2).
\end{equation}

From this equation, it follows that $\theta$ cannot be $\pi$ for that would
yield the condition: $1+a+b \le 0.$ However, the remaining $n$ possible
choices for $\theta$ yield allowed regions which are bounded by the
hyperbola: $1+a+b = \cot^2(\theta/2).$ The hyperbola is symmetric about $a=b$
and has asymptotes at $a=\tan^2(\theta/2)$ and $b=\tan^2(\theta/2).$ For
example, if n=1, we are dealing with (32) which is the symmetric curve of
the Lieber diagram. In that case, $\theta=\pi/3$ and our analysis gives the
allowed region described by: $3ab-a-b-1 \ge 0.$ To ascertain which $\theta$
is the physical one, we insist that the time duration between collisions are
all positive. The formulas for the time ratios here are:
\begin{eqnarray}
t_{2j}/t_{2j-1}&=& -\cos((j-1/2)\theta)/\cos((j+1/2)\theta)\mbox{ and} \\
t_{2j+1}/t_{2j}&=&-\sin(j\theta)/\sin((j+1)\theta).
\end{eqnarray}
These must be nonnegative for $j\le n-1,$ which requirement holds only for $(n-1)\pi/n\le\theta\le\pi.$ The only suitable $\theta$ from Eq.~(\ref{thetaqu}) is $(2n-1)\pi/(2n+1)$ and so the allowed region for $(31)^n$ is given by $ab\cot^2((2n-1)\pi/(2(2n+1)))-a-b-1 \ge 0.$  That is, it lies to the right of the right branch of the hyperbola given by:
\begin{equation}
ab\cot^2\left(\frac{(2n-1)\pi}{2(2n+1)}\right)-a-b-1 = 0.
\end{equation}

\newpage
\section{Conclusions}
We have shown that there exist regions in the Lieber diagram for which the processes $(2)(32)^n,$ $(32)^n,$ $(21)^n,$ $(21)^n(2)$ and $(31)^n$ can occur classically
and that, except for the last type, these regions are non-overlapping.  If mass ratios are
chosen in figure 3 which lie in the forbidden region, there is a unique process whereby particle 1 can capture particle 2.  (The forbidden region excludes processes of the fifth type.)

\appendix
\section{Proof of Lemma 1}
We will first demonstrate the impossibility of the sequence (2312).  In what follows, the respective velocities of 1, 2 and 3 will be given in column vector form with $\vec{v}$ representing the three velocities and $\vec{r}$ representing the three positions.  Let the velocities and positions for 1, 2 and 3 immediately after the first collision be
\begin{equation}
\vec{v}=\left(\begin{array}{c}0\\v\\w e^{i\alpha}\end{array}\right)
\hspace{1 cm}\mbox{ and }\hspace{1 cm}
\vec{r}=\left(\begin{array}{c}0\\-r\\0\end{array}\right).
\end{equation}
The quantities $v$, $w$, and $r$ are positive reals and $\alpha$ is a real angle.  We are working in the reference frame of particle 1.  

In sequences of the form (232), the scattering angle of the second collision is given by $\theta=\pi+2\alpha$ and in sequences of the form (231), the scattering angle is given by $\theta=\alpha+\sin^{-1}((\sin\alpha)/a)$.

The time until the next collision is given by $t_1=r/v$.  At the end of this interval, the positions are:\begin{equation}
\vec{r}=\left(\begin{array}{c}0\\0\\rw e^{i\alpha}/v\end{array}\right).
\end{equation}
Let the scattering angle of the second collision be $\theta$.  Then after that collision the velocities are given by
\begin{equation}
\vec{v}=\left(\begin{array}{c}v(1-e^{i\theta})/(1+a)\\v(1+ae^{i\theta})/(1+a)\\w e^{i\alpha}\end{array}\right).
\end{equation}

or in the reference frame of 2:
\begin{equation}
\vec{v}=\left(\begin{array}{c}-v e^{i\theta}\\0\\w e^{i\alpha}-v(1+a e^{i\theta})/(1+a)  \end{array}\right).
\end{equation}

The time until the third collision is given by
\begin{equation}
t_2=\frac{rw e^{i\alpha}/v}{(1+a e^{i\theta})v/(1+a)-
w e^{i\alpha}}.
\end{equation}
In order for $t_2$ to be real and positive, $(1+a e^{i\theta})v/(1+a)$ must have the same phase as and be greater in magnitude than $w e^{i\alpha}$.  Assuming this to be the case, $t_2$ can be expressed as:
\begin{equation}
t_2=\frac{rw/v}{(\sqrt{1+2a\cos(\theta)+a^2})v/(1+a)-w}.
\end{equation}
At the end of this interval the positions are
\begin{equation}
\vec{r}=\left(\begin{array}{c}-v e^{i\theta} t_2\\0\\0\end{array}\right).
\end{equation}

Let the third collision have scattering angle $\phi$.  Then after the third collision the velocities are given by
\begin{equation}
\vec{v}=\left(\begin{array}{c}-v e^{i\theta}\\ b(1 - e^{i\phi})(w e^{i\alpha}(1+a)-v(1+ae^{i\theta}))/(1+a)(1+b)\\(b + e^{i\phi})(w e^{i\alpha}(1+a)-v(1+ae^{i\theta}))/(1+a)(1+b)\end{array}\right).
\end{equation}
The time until the fourth collision is given by
\begin{equation}
t_3=-\frac{ve^{i\theta}t_2}{(b + e^{i\phi})(w e^{i\alpha}(1+a)-v(1+ae^{i\theta}))/(1+a)(1+b)+ve^{i\theta}}.
\end{equation}  In order for $t_3$ to be positive, $(b+e^{i\phi})(w e^{i\alpha}(1+a)-v(1+ae^{i\theta}))/(1+a)(1+b)$ must have the same phase as and be greater in magnitude than $-ve^{i\theta}$.  The magnitude of the former is $\sqrt{b^2+2b\cos{\phi}+1}(v\sqrt{a^2+2a\cos{\theta}+1}/(1+a)-w)/(1+b)$.  The factor $v\sqrt{a^2+2a\cos{\theta}+1}/(1+a)-w$ has a value less than $v$ and the factor $\sqrt{b^2+2b\cos{\phi}+1}/(1+b)$ is less than $1$ so the condition for the third time interval to be positive cannot be met.  Therefore the proposed collision sequence is impossible.

We will next demonstrate the impossibility of processes of the form (32313).  

Immediately after the first collision, in the frame of 1, 2 is moving away from 1 and 3 is moving toward 1.  Let the velocity of 3 be $w$ and the velocity of 2 be $v\exp(i\alpha).$  We view the next collision, (2), in the frame of particle 1.  After the collision (2), 2 needs to have its velocity shifted by a phase of $\pi$ so that it changes from an outgoing particle to an ingoing particle.  The collision matrix for collision (2) is:
\begin{equation}
\left(
\begin{array}{cc}
1 & \frac{2b}{a+b}\sin(\theta/2)\exp[i(\theta+\pi)/2] \\
0&\exp(i\theta)
\end{array}
\right)
\end{equation}
Before the second collision, the velocities of 2 and 3 are $(v\exp(i\alpha), w)^T$.  If we apply the above matrix to this vector, we find the velocities after the second collision:  
\begin{equation}
(v\exp(i\alpha)+2bw\sin(\theta/2)\exp[i(\theta+\pi)/2]/(a+b),w\exp(i\theta))^T.\label{aftersecond}
\end{equation}

As mentioned, we need to have the velocity of 2 shifted by a phase of $\pi$.  This requires that the term $2bw\sin(\theta/2)\exp[i(\theta+\pi)/2]/(a+b)$ have opposite phase from $v\exp(i\alpha)$ and be larger in magnitude than $v$.  For the phase of the correction term to be right, we need $\theta=2\alpha+\pi$.  Plugging this into Eq.~\ref{aftersecond} gives the velocities as:
\begin{equation}
((v-2bw\cos\alpha/(a+b))\exp(i\alpha), -w\exp(2i\alpha))^T
\end{equation}
and for the velocity of particle 2 to be turned around, we need:

\begin{equation}
v<\frac{2bw\cos\alpha}{a+b}.
\end{equation}

We will analyze the third collision by working in the frame of 1 before the collision and by working in the frame of 2 after the collision.  That way, before the collision, 3 is moving away from 1 and 2 is moving toward 1 and after the collision, 1 is moving away from 2 and 3 is moving toward 2.  At the moment of the third collision, 2 replaces 1 at the origin.  Thus before the collision, 3 is moving away from the origin and after the collision it is moving toward the origin.  So we need the collision to change the phase of particle 3 by $\pi$.
The matrix for collision (3) in which we work in the frame of 1 before the collision and the frame of 2 after the collision is:
\begin{equation}
\left(
\begin{array}{cc}
-\exp(i\phi)&0\\
-(1+a\exp(i\phi))/(1+a)&1
\end{array}
\right).\label{231}
\end{equation}

Before the the third collision, the velocities of 2 and 3 are :
\begin{equation}
\exp(i\alpha)((v-2bw\cos\alpha/(a+b)), -w\exp(i\alpha))^T
\end{equation} which may also be written as:
\begin{equation}
-\exp(i\alpha)(v', w\exp(i\alpha))^T
\end{equation}
where $v'$ is a positive number.  From the matrix \ref{231}, the velocity of 3 after the collision is $-\exp(i\alpha)(w\exp(i\alpha)-(1+a\exp(i\phi))v'/(1+a))$ the phase of which must differ from that of $w\exp(i\alpha)$ by 
$\pi$.  This requires that $(1+a\exp(i\phi))v'/(1+a)$ must have the same phase as and be greater in magnitude than $w\exp(i\alpha)$.  For the latter to have the same phase as the former, we need 
\begin{equation}
\phi=\alpha+\sin^{-1}\left(\frac{\sin\alpha}{a}\right).
\end{equation}
If the above value for $\phi$ is used, the magnitude of $(1+a\exp(i\phi))v'/(1+a)$ can be shown to be $(\cos\alpha+\sqrt{a^2-\sin^2\alpha})v'/(1+a)$.  Since this must be greater than $w$, the condition for the fourth collision is:
\begin{equation}
\left[\frac{2bw\cos\alpha}{a+b}-v\right]\frac{(\cos\alpha+\sqrt{a^2-\sin^2\alpha})}{1+a}>w
\end{equation}
After the third collision, in the reference frame of 2, the velocities of 1 and 3 are:
\begin{equation}
-\exp(2i\alpha)(-v'(\sqrt{a^2-\sin^2\alpha}+i\sin\alpha)/a, w-(\cos\alpha+\sqrt{a^2-\sin^2\alpha})v'/(1+a))^T.
\end{equation} or
\begin{equation}
-\exp(2i\alpha)(-v'\exp(i(\phi-\alpha)), -w')^T
\end{equation}
where $w'$ is a positive number.
We can analyze the fourth collision in the reference frame of particle 2.  Before the fourth collision, 1 is moving away from 2 and after, 1 is moving toward 2.  Thus the fourth collision must change the phase of the velocity of 1 by $\pi$.  The matrix which mediates this collision is:
\begin{equation}
\left(
\begin{array}{cc}
1&2b\exp[i(\psi+\pi)/2]\sin(\psi/2)/(1+b)\\
0&\exp(i\psi)
\end{array}
\right).\label{313}
\end{equation}

Before the fourth collision, the phase of the velocity of 1 minus the phase of the velocity of 3 is $\phi-\alpha$.  Thus the scattering angle of the fourth collision must be $2(\phi-\alpha)+\pi$ and the matrix may be rewritten as:
\begin{equation}
\left(
\begin{array}{cc}
1&-2b\exp[i(\phi-\alpha)]\cos(\phi-\alpha)/(1+b)\\
0&-\exp(2i(\phi-\alpha))
\end{array}
\right).
\end{equation}
The velocity of 1 after the fourth collision may be written as:
\begin{equation}
-\exp(2i\alpha)(-v'\exp(i(\phi-\alpha))+2bw'\exp[i(\phi-\alpha)]\cos(\phi-\alpha)/(1+b))
\end{equation}
or
$\exp(i(\phi+\alpha))(v'-2bw'\cos(\phi-\alpha)/(1+b))$
whereas before it was $\exp(i(\phi+\alpha))v'$.  Thus the condition for the fifth collision is:
\begin{equation}
2bw'\cos(\phi-\alpha)/(1+b)>v'
\end{equation} which can be written in full as:
\begin{equation}
\frac{b}{1+b}\left[-w+\left(\frac{2bw\cos\alpha}{a+b}-v\right)\frac{(\cos\alpha+\sqrt{a^2-\sin^2\alpha})}{1+a}\right]\frac{2\sqrt{a^2-\sin^2\alpha}}{a}>\frac{2bw\cos\alpha}{a+b}-v.
\end{equation}
The factors $(\sqrt{a^2-\sin^2\alpha})/a$ and $(\cos\alpha+\sqrt{a^2-\sin^2\alpha})/(1+a)$ are less than or equal to 1 so we can replace them with 1 and the inequality for the fifth condition remains true:
\begin{equation}
\frac{2b}{1+b}\left(\frac{2bw\cos\alpha}{a+b}-v-w\right)>\frac{2bw\cos\alpha}{a+b}-v.
\end{equation}
The following sequence of algebraic manipulations demonstrates the impossibility of the sequence (32313):
\begin{eqnarray}
&&\frac{2b}{1+b}\left(\frac{2bw\cos\alpha}{a+b}-v-w\right)>\frac{2bw\cos\alpha}{a+b}-v\\&\Rightarrow&\frac{2bw\cos\alpha}{a+b}\left(\frac{b-1}{b+1}\right)>\left(\frac{b-1}{b+1}\right)v+\frac{2bw}{1+b}\\
&\Rightarrow&\frac{2bw}{1+b}\left[\frac{(b-1)\cos\alpha}{a+b}-1\right]
>\left(\frac{b-1}{b+1}\right)v>0\\
&\Rightarrow& (b-1)\cos\alpha > a+b 
\end{eqnarray}
which is a contradiction.



\section{Proof of Lemma 2}
Suppose particles 1 and 2 collide and that before the collision, the
velocities of 2 and 3 are $p=p_0 \exp(i\phi)$ and $q=q_0 \exp(i\psi).$ Then
after the collision, their velocities, $p^{\prime}$ and $q^{\prime},$ are
given by:
\begin{equation}
\left(
\begin{array}{c}
p^{\prime} \\
q^{\prime}
\end{array}
\right)= \left(
\begin{array}{cc}
x & 0 \\
(x-1)/(a+1) & 1
\end{array}
\right) \left(
\begin{array}{c}
p \\
q
\end{array}
\right)= \left(
\begin{array}{c}
xp \\
q+p(x-1)/(a+1)
\end{array}
\right)
\end{equation}
After 1 and 2 collide, 1 and 3 must collide if we are considering a process
of the form $(32)^n$ or $(2)(32)^n.$ Now if the previous collision between 1
and 2 was preceded by one between 1 and 3, 3 is moving radially away from 1
at the moment of the collision between 1 and 2. So for another collision to
occur between 1 and 3, the velocity of 3 after the collision, $q^{\prime}=q+p(x-1)/(a+1),$ must be directed toward the origin and so be
antiparallel to its velocity before the collision, $q.$ But if $q$ and $q+p(x-1)/(a+1)$ are antiparallel, the same is true of $q$ and $p(x-1)/(a+1).$
The phase of $x$ is $\alpha$ and by a trigonometric identity, the phase of $x-1$ is $(\alpha+\pi)/2.$ Therefore, for $q$ and $p(x-1)/(a+1)$ to be
antiparallel, we need:
\begin{eqnarray}
\psi+\pi=\phi+(\alpha+\pi)/2&\Rightarrow& \\
\alpha=2(\psi-\phi)+\pi &&
\end{eqnarray}
The above derivation was for the required scattering angle of a collision
between 1 and 2 in order for a collision between 1 and 3 to follow. But
since the final result is independent of any mass ratios, it also works for
the case of a collision between 1 and 3. However for that case, by $\phi$ we
mean the angle of incidence of particle 3 and by $\psi$ we mean the initial
angle of 2's velocity.

Following the collision between 1 and 2 described above, the phase of 2's
velocity is: $\psi^{\prime}=\phi+\alpha=2\psi-\phi+\pi$ and the phase of 3's
is $\phi^{\prime}=\psi+\pi.$ Now if these are plugged into the formula for
scattering angle, we obtain:
\begin{equation}
\alpha^{\prime}=2(\psi^{\prime}-\phi^{\prime})+\pi =2(\psi-\phi)+\pi=\alpha.
\end{equation}
This means that if four collisions occur in a row (either (2323) or (3232))
the second and third scattering angles are equal. The first and fourth
collisions are required because the collisions discussed above were assumed
to be preceded and followed by collisions. Processes which lead to capture
leave 1 and 2 with the same velocity. Two particles with the same velocity
may be said to collide after an infinite time so the process concludes with
a virtual collision (3). Since this may be counted as the fourth
collision, all collisions have the same scattering angle with the exception
of the first.

\end{document}